\documentclass{aastex}
\usepackage{emulateapj5}

\newcommand{\ha}{H$\alpha$}
\newcommand{\hb}{H$\beta$}

\newcommand{\kms}{km\,s$^{-1}$}

\shortauthors{HALL}
\shorttitle{FAST MOVING L DWARF} 

\begin{document}

\journalinfo{Accepted to ApJL October 15, 2002} 
\submitted{}
\title{2MASS 1315$-$2649:
A High Proper Motion L Dwarf with Strong H$\alpha$ Emission\altaffilmark{1}}
\altaffiltext{1}{
Based on observations collected at the European Southern Observatory (ESO), 
Chile, for proposal \#69.A-0068.}

\author{Patrick B. Hall
}
\affil{
Pontificia Universidad Cat\'{o}lica de Chile, Departamento de Astronom\'{\i}a
y Astrof\'{\i}sica, 
Casilla 306, Santiago 22, Chile,
and Princeton University Observatory, Princeton, NJ 08544-1001;
E-mail: phall@astro.puc.cl
}

\begin{abstract}
In \markcite{hal02lha}{Hall} (2002) I reported that 2MASSI J1315309$-$264951
is an L dwarf with strong H$\alpha$ emission.
Two spectroscopic epochs appeared to show that the H$\alpha$ was variable,
decreasing from 121\,\AA\ to 25\,\AA\ EW,
which I interpreted as a flare during the first observation.
\markcite{giz02}{Gizis} (2002) independently discovered this object, and his 
intermediate spectroscopic epoch shows H$\alpha$ with 97\,\AA\ EW.
A new fourth epoch of spectroscopy again shows a very large H$\alpha$ EW
(124\,\AA), confirming this object to be a persistent, strong H$\alpha$ emitter.
Whether the \ha\ is steady (like 2MASS 1237+6526) or from continuous strong
flaring (like PC0025+0447) remains unclear.  Imaging confirms that 
2MASS\,1315$-$2649 has a high proper motion (0\farcs71 year$^{-1}$), 
corresponding to a transverse velocity of $\sim$76\,\kms\ at its distance of
$\sim$23 pc.  Thus 2MASS\,1315$-$2649 is consistent with being $\gtrsim$2~Gyr 
old and therefore relatively massive.  If that is so, the correlation of
\ha\ activity with mass found by \markcite{giz00}{Gizis} {et~al.} (2000) 
would seem to support the continuous strong flaring scenario,
though it does not rule out a brown dwarf binary accretion scenario.
\end{abstract}


\section{Introduction}  \label{INTRO}

M dwarf stars often show H$\alpha$ in emission,
but the frequency of \ha\ emission peaks around type M7 and declines for
later-type L and T dwarfs \markcite{giz00}({Gizis} {et~al.} 2000). 
The only objects of type L5 or later seen to exhibit strong \ha\ are 
the persistent \ha-emitting T6.5 dwarf 2MASSI J1237392+652615 
(\markcite{bur02}{Burgasser} {et~al.} 2002; hereafter B02), the flaring L5
dwarf 2MASSI J0144353$-$071614 \markcite{lie02}({Liebert} {et~al.} 2003) 
and the L5 dwarf 2MASSI J1315309$-$264951 
(\markcite{hal02lha,giz02}{Hall} 2002; {Gizis} 2002; hereafter H02 and G02). 
I report on a new spectrum and a proper motion estimate 
of this latter object, hereafter 2MASS 1315$-$2649, which are of interest
given the rarity of L and T dwarfs with strong \ha\ and 
the uncertain mechanism(s) driving such emission.

\section{Spectroscopy}  \label{SPEC}

Spectroscopy of 2MASS\,1315$-$2649 was secured beginning at 00:15
on UT 2002 September 9 using the ESO Multi-Mode Instrument (EMMI)
at the ESO 3.6m New Technology Telescope (NTT).
A 316 lines/mm grating blazed at 6200\,\AA\ was used to cover the wavelength
range 4500$-$7000\,\AA\ at 1.575\,\AA/pixel and resolution 4.6\,\AA, given
the 1\arcsec\ slit.  Three 5-minute exposures were obtained at
the parallactic angle.  Despite cirrus and the high airmass.
the unresolved \ha\ line is readily visible on the individual exposures.
The sky-subtracted spectrum was extracted using an aperture eight pixels
(2\farcs664) wide.  CD$-$32\arcdeg9927 \markcite{ham94}({Hamuy} {et~al.} 1994)
was used for flux calibration.

As in H02, the continuum was measured at 6400$-$6550\,\AA\ 
and 6575$-$6725\,\AA, and the average used for the equivalent width calculation.
I find an \ha\ EW=124$\pm$55\,\AA\
($\log(L_{\rm H\alpha}/L_{\rm bol})\simeq -4.01$, following H02),
compared with 121$\pm$31\,\AA\ in March 2001 and 25$\pm$10\,\AA\ in August 2001
(H02), and 97\,\AA\ in May 2001 (G02). 
Either the August 2001 measurement is a spurious outlier and the object has
persistent, nonvariable \ha\ emission of EW$\simeq$100\,\AA\ 
($\log(L_{\rm H\alpha}/L_{\rm bol})\simeq -4$), or the \ha\ emission
is continuously flaring or slowly variable with that same EW.
Smaller uncertainties on the \ha\ fluxes than the current 25-40\%
are needed to distinguish between these scenarios.

Note that \hb\ was not detected to a 3$\sigma$ limiting flux equal to 37\% of
the \ha\ flux.  That is, the Balmer decrement is \ha/\hb$>$2.7 in this object.
This is larger than the value 2.3 measured for 
PC0025+0447 by \markcite{mea94}{Mould} {et~al.} (1994),
but similar to the limit \ha/\hb$>$4.2 measured for 2MASS 1237+6526 by
\markcite{bur00}{Burgasser} {et~al.} (2000); see \S\ref{DISC}.

\section{Astrometry}  \label{ASTRO}

An unfiltered CCD image was taken just prior to the NTT spectroscopy.  The
exposure time was 3 minutes and the pixel scale was 0\farcs333$\pm$0\farcs001.
The image shows both 2MASS\,1315$-$2649 and
USNO\,J131531.230$-$264953.01, an optically brighter star slightly 
south of east.  The separation between the objects (epoch 2002.685) is
7\farcs14$\pm$0\farcs05, including the systematic uncertainty due to 
differential refraction, at PA 96\fdg2 East of North.
This is significantly different from the 2MASS-USNO coordinate separation of
4\farcs23$\pm$0\farcs32 at PA 114\fdg5, where uncertainties of 0\farcs25
and 0\farcs2 have been assumed for the USNO (mean epoch 1977.792) and 
2MASS (epoch 1998.411) positions, respectively.
Systematic errors between USNO and 2MASS are possible, but an $R=12.6$ star
located only 38\arcsec\ distant has coordinates that differ by only 0\farcs23.
Thus it appears that the objects' separation has increased significantly since
the 2MASS observations, as suggested by H02. 

Assuming the USNO star has negligible proper motion, 
2MASS\,1315$-$2649 has a proper motion of 
0.71$\pm$0.07 arcsec year$^{-1}$ at PA 253\fdg8.\footnote{
The predicted E-W separation of the two objects at the time of the H02 CTIO
observations was 5\farcs3, vs. the observed 5\farcs5$\pm$0\farcs3.}
This is approximately ten times larger than expected if
2MASS\,1315$-$2649 is a member of the TW Hya association of young stars
(it was selected for observation by G02 as a candidate member).
At the distance of $\sim$23$\pm$2 pc estimated (using Fig. 7.21 of
\markcite{bur01}{Burgasser} 2002) from its apparent magnitude
and spectral type (L5, from the high quality spectrum of G02), 
the proper motion corresponds to a transverse velocity of
$\sim$76$\pm$10\,\kms.  This rather large velocity suggests that
2MASS\,1315$-$2649 is relatively old and massive, as discussed further below. 

\section{Discussion}  \label{DISC}

Four epochs of spectroscopy now exist for 2MASS 1315$-$2649, 
three of which show $\log(L_{\rm H\alpha}/L_{\rm bol}) \simeq -4$.
Thus this object is a persistent, strong \ha\ emitter, of which only a few other
examples are known among the M, L and T dwarfs: the M9.5 dwarf PC 0025+0447
(\markcite{mbz99}{Mart\'{\i}n}, {Basri}, \& {Zapatero  Osorio} 1999; 
hereafter MBZ99), the T6.5 dwarf 2MASS 1237+6526 (\S\,\ref{INTRO}) and possibly 
the M9 dwarf LP 412-31 (\markcite{giz00}{Gizis} {et~al.} 2000, \S\,6.2), which 
has just two spectral epochs and is not discussed further.

The M dwarf PC0025+0447 has always been observed to have strong \ha\ emission
($\log(L_{\rm H\alpha}/L_{\rm bol}) \simeq -3.4$),
yet it also shows classic signatures of flaring.
MBZ99 show that the \ha\ strength varies on timescales of days to years
by up to a factor of four (from EW 390$\pm$45\,\AA\ to 110$\pm$15\,\AA), 
\ion{He}{1} emission is sometimes seen, and the optical spectrum is veiled by 
an additional blue continuum  when the emission lines are stronger.
A T Tauri interpretation is disfavored by the lack of an obvious near-infrared
excess and by line profiles dissimilar from those of T Tauri stars.
Thus MBZ99 conclude that PC0025+0447 has a highly active chromosphere and/or
corona; it very likely has continuous strong flares.  Continuous weak flaring
occurs in many late-type dwarfs \markcite{tks00}({Tsikoudi} {et~al.} 2000),
but typically with \ha\ equivalent widths ten to a hundred times smaller.

The T dwarf 2MASS 1237+6526 has \ha\ emission of
$\log(L_{\rm H\alpha}/L_{\rm bol}) = -4.3$ (B02) 
that is constant to within 10\%\ over three epochs spanning two years.
Thus it is unlikely to have continuous strong flares.  B02 suggest that it 
could be a young, very low-mass brown dwarf still accreting material from a 
surrounding disk, similar to the weak-lined T Tauri stars. 
However, they cannot completely rule out 
a close brown dwarf binary accretion hypothesis.

Existing observations of 2MASS\,1315$-$2649 are insufficient to determine to
which of the above two objects it is most similar.\footnote{According to the
NASA/IPAC Extragalactic Database, 2MASS\,1315$-$2649 was not detected in the
radio by NVSS nor in X-rays by ROSAT.} Three of four spectra
taken over 1.5 years are consistent with a constant \ha\ flux (to within
$\sim$30\%).  The discrepant observation is of the lowest
quality, but other than that there is no a priori reason to discard it.  

However, the proper motion and Balmer decrement of 2MASS 1315$-$2649
may be clues to the origin of its \ha.

Since the velocity dispersion of disk stars increases with age,
the large transverse velocity inferred for 2MASS\,1315$-$2649 from its
proper motion suggests that it is $\gtrsim$2~Gyr 
old \markcite{dea02}({Dahn} {et~al.} 2002) and therefore relatively massive for
its spectral type.  This is consistent with the correlation noted for weaker
\ha\ emitters by \markcite{giz00}{Gizis} {et~al.} (2000), namely that 
``active L dwarfs are drawn from an older, more massive population."
Unless the incidence of close 
brown dwarf binaries is a strong function of mass, strong magnetic fields 
appear to be the only plausible explanation for this trend.  
The inferred high mass of 2MASS\,1315$-$2649 thus suggests it is a
continuously flaring object with an unusually strong magnetic field.

Gas that is optically thin in the Balmer lines will have \ha/\hb$\lesssim$3.3
for all 2500$<$$T$$<$$2\times10^6$\,K \markcite{fer80,mar88}({Ferland} 1980;
{Martin} 1988), whereas Balmer decrements as high as $\sim$5 are seen in 
T Tauri stars, produced in optically thick, low-ionization 
gas at $T$$\sim$$6500$\,K \markcite{lam96}({Lamzin} {et~al.} 1996).
These observations appear 
nicely consistent with PC0025+0447 having \ha/\hb$\sim$2.3 due to flares
and 2MASS 1237+6526 having \ha/\hb$>$4.2 due to disk (or close binary)
accretion, but this is misleading.  Flares are complex processes arising in 
chromospheric gas with $n_e$$\sim$10$^{12}$\,cm$^{-3}$,
sufficient to be optically thick in lower-order Balmer lines
\markcite{hea91}({Houdebine} {et~al.} 1991), and so a chromospheric origin for 
emission with \ha/\hb$>$3.3 cannot be ruled out.
In fact, dwarfs later than M4 often have \hb/\hb$>$3.3, and values up to 15
have been seen \markcite{grh02}({Gizis}, {Reid}, \& {Hawley} 2002).
Given this large range of \ha/\hb\ seen in normal dwarf star chromospheres,
2MASS\,1315$-$2649 could have continuous strong flares like PC0025+0047
despite having a larger Balmer decrement (\ha/\hb$>$2.7 at 3$\sigma$).

\section{Conclusion}  \label{CONCL}

The origin of the persistent, strong \ha\ emission of
2MASS 1315$-$2649 remains unclear. 
Its inferred old age makes it unlikely to be a brown dwarf T Tauri analog
and its inferred high mass suggests that it has continuous strong flares,
but existing observations cannot rule out an accreting brown dwarf binary 
scenario.  As with the other two known persistent, strong \ha\ emitting
	dwarf stars, determining the origin of the emission 
	will require high-quality spectroscopic or photometric monitoring.

\acknowledgements

I thank S. Ellison, G. Ferland and the referee, and Fund.  Andes 
and FONDECYT grant 1010981 for financing. 

\bibliography{}


\end{document}